\documentclass[twocolumn,secnumarabic, amssymb, superscriptaddress, nobibnotes, aps, pra]{revtex4-2}

\usepackage{graphicx}
\usepackage{dcolumn}
\usepackage{bm}
\usepackage{amsmath,amssymb,mathtools}
\usepackage{stmaryrd}
\usepackage{times}
\usepackage[euler]{textgreek}
\usepackage[bbgreekl]{mathbbol}
\usepackage{graphicx}
\usepackage{bm,color}
\usepackage{natbib,hyperref}
\usepackage[capitalize]{cleveref}
\usepackage{multirow}

\hypersetup{colorlinks=true,linkcolor=blue,citecolor=blue,urlcolor=blue}

\begin{document}
	
	\title{Staircase mechanical energy growth in optomechanical systems of median mechanical frequencies}
	\author{Yi Xiao }
	\thanks{These authors contributed equally to this work.}
	\affiliation{Fujian Key Laboratory of Light Propagation and Transformation \& Institute of Systems Science, 
		College of Information Science and Engineering, Huaqiao University, Xiamen 361021, China}
		\author{Yi Wu }
		\thanks{These authors contributed equally to this work.}
	\affiliation{Shanghai United Imaging Healthcare Co., Ltd., 2258 Chengbei Rd, Jiading District, Shanghai 201807, China}
		\author{Qi-Kai Zhan}
	\affiliation{Fujian Key Laboratory of Light Propagation and Transformation \& Institute of Systems Science, 
		College of Information Science and Engineering, Huaqiao University, Xiamen 361021, China}
	\author{Jin Lian Zhang}
	\affiliation{Fujian Key Laboratory of Light Propagation and Transformation \& Institute of Systems Science, 
		College of Information Science and Engineering, Huaqiao University, Xiamen 361021, China}
	\author{Bing He}
	\email{bing.he@umayor.cl}
	\affiliation{Multidisciplinary Center for Physics, Universidad Mayor, Camino La Pir\'{a}mide 5750, Huechuraba, Chile}
	\author{Qing Lin}
	\email{qlin@hqu.edu.cn}
	\affiliation{Fujian Key Laboratory of Light Propagation and Transformation \& Institute of Systems Science, 
		College of Information Science and Engineering, Huaqiao University, Xiamen 361021, China}
	
	\begin{abstract}
	  Owing to the radiation-force-induced nonlinearity, cavity optomechanical systems (COMS) exhibit rich 
	  dynamical phenomena such as back-action induced oscillation, chaos, mechanical amplitude locking, and anomalous stabilization, which occur under different driving conditions and different system parameters. 
	  We here identify a previously unknown dynamical pattern of staircase evolution for the energy of mechanical resonator, when a COMS with neither very large nor very small built-in mechanical frequency is driven by a two-tone field, which satisfies a condition that the frequency difference of the two tones matches the built-in mechanical frequency. The properties of this phenomenon are analyzed for the different system parameters due to fabrication such as mechanical frequencies and quality factors, as well as under the varied driving conditions such as unequal drive tone powers and mismatched drive tone difference from the mechanical frequency.
	  Some special features, such as an emergent bifurcation due to the tone power difference, together with the 
	  totally different responses of the system to the drive tone mismatches of opposite signs, are discovered to exist only in this type of COMS with median mechanical frequencies. This work fills a gap in the study of the dynamics of COMS under two-tone drives. In the aspect of applications, the rapid increase of mechanical energy exhibited in the phenomenon promises phonon laser generation, and the sensitive dynamical response to the drive tone mismatches offers a potential approach to high-precision sensing.
	\end{abstract}
	
	\maketitle
	\section{Introduction}
	COMS is a type of nonlinear dynamical systems with their nonlinearity originating from the interaction between the associated intracavity field and mechanical resonator. Radiation force from the momentum of photons induces a displacement $x_m(t)$ of mechanical resonator, such as a movable mirror in Fabry-P\'erot cavities or the boundaries of microtoroid cavities, microsphere cavities, leading to a variation of the lengths or radii of the optical cavities. This in turn modifies the resonance frequency of the optical cavity and gives rise to a nonlinear response to the pump field \cite{OMS}. Since the radiation force is proportional to the intracavity field intensity or the associated photon number $|a|^2$, such optomechanical interaction can be enhanced by increasing the power of pump field. As a result, COMS can exhibit strong nonlinearity due to a light-matter interaction, enabling the applications in many areas such as optomechanically induced transparency \cite{OMIT1,OMIT2,OMIT3}, optomechanically induced amplification \cite{OMIA1,OMIA2,OMIA3}, optomechanically induced nonreciprocity \cite{nonre1,nonre2}, phonon laser \cite{phonon01,phonon1, phonon11, phonon2, phonon3, phonon33,pl}, optoemchanical frequency combs \cite{ofc1, ofc2, ofc22, ofc3, ofc, comb-s}, mass sensing \cite{ms01, ms02}, and force sensing \cite{fs1,fs2,fs3,fs4,fs5}.
	
	Rather than the nonlinear aspect, the early-day research on COMS mainly focused on quantum effects. 
	Through a beamsplitter-type coupling induced by red-detuned single-tone drives, mechanical resonators can be cooled down close to ground-state \cite{cooling1, cooling2,cooling3, cooling4,cooling5, cooling6, cooling7, cooling8}. On the other hand, a squeezing-type coupling from blue-detuned drives can realize the entanglement between intracavity field and mechanical resonator \cite{entanglement1, entanglement2, entanglement20, entanglement3, entanglement4, entanglement4b, entanglement5, entanglement6, entanglement7}. Beyond these linearized scenarios, significant nonlinear optomechanical effects were believed to exist under either strong single-photon coupling strength 
	or higher drive power. One example is that when the drive power exceeds a certain threshold, a Hopf bifurcation will occur in COMS, so that the mechanical resonator undergoes a transition from steady state to oscillation \cite{ss1,ss2, ss3,ss4,ss5,ss5b,ss6,ss7,ss8,ss80,ss9,ss91,ss92,ss10}. However, even in the simplest situation of single-tone laser drive, the realistic behaviors of COMS are much more complex than expected. For example, near the point where the drive is blue-detuned by the built-in mechanical frequency, a COMS can enter a dynamical evolution of anomalous stabilization within a certain range of drive powers, and the mechanical energy will increase in a step-like manner \cite{as}.

Similar dynamical behaviors can occur more easily when a COMS is driven by a two-tone field. It is found that the mechanical energy increment of a COMS fabricated in the unresolved sideband regime (the built-in mechanical frequency $\omega_m$ is less than the cavity damping rate $\kappa$) also exhibits a step-like increment pattern, when it is driven by a two-tone field with the tone difference matching the built-in mechanical frequency (it is regarded as a resonance point) \cite{nr}. In this case, each step of mechanical energy lasts for only one oscillation period before the pulsed intracavity field pushes it to a higher step, so that the mechanical energy steps exhibits a highly regular pattern in a special nonlinear resonance. On the other hand, if the built-in mechanical frequency becomes sufficiently higher than the cavity damping rate, e.g., $\omega_m=10\kappa$, the mechanical oscillation of the system will be totally locked into a series of discrete orbits, on which the three elements of oscillation, the amplitude, frequency and phase can be completely frozen \cite{lo1, lo2}. The difference between the step-like energy increase pattern and the frozen orbits is only in the associated ratios $\omega_m/\kappa$. This observation naturally raises this question: how will the dynamical behaviors gradually transit from those of a nonlinear resonance exhibiting the regular mechanical energy steps to those of the locked orbits, as the ratio continuously varies from $\omega_m/\kappa<1$ to $\omega_m/\kappa\gg 1$ (especially the behaviors between the two extremes)? Here we investigate the dynamical behaviors of the system when the ratio $\omega_m/\kappa$ is in the intermediate range. We find that within this intermediate range, the mechanical oscillator has a different step-like evolution pattern, which exhibits both oscillation and ever changing step length for the mechanical energy. The discovery of this pattern completes the whole picture about how the built-in mechanical frequency determines the optomechanical dynamics under a frequency-matched two-tone field.

Various applications are possible based on the OMS driven by two-tone field. The nonlinear resonance phenomenon in the unresolved sideband regime with $\omega_m/\kappa<1$ \cite{nr}, which also leads to pulsed intracavity field, can effectively broaden the bandwidth of generated optical frequency comb \cite{ofc}. The phenomenon of locked orbits, which completely locks the amplitude of mechanical resonator \cite{lo1, lo2}, can make the system become highly sensitive to external perturbations, such as a change in mechanical frequency due to nano-particle adsorption, the variations of mechanical oscillator displacement induced by weak forces, and others, enabling ultra-high-precision sensing of mass \cite{ms1, ms2} and tiny forces \cite{wf1, wf2}. On the other hand, the oscillatory staircase phenomenon studied in this work can excite relatively large mechanical oscillation under lower driving power, potentially beneficial to the development of phonon lasers \cite{phonon01,phonon1, phonon11, phonon2, phonon3, phonon33,pl}. The regime of intermediate values near $\omega_m/\kappa=1$ was predicted as the optimum for generating mechanic-optical frequency comb under single-tone drive \cite{comb-s}. For the same reason the corresponding properties due to two-tone drives should be well understood too.

	The rest of the paper is organized as follows. Sec. \ref{sec2} details the setup, dynamics, and the associated notation for the work, together with an illustration how step-like evolution arise under a single-tone drive. The  concerned phenomenon of oscillatory staircase energy evolution are introduced in Sec. \ref{sec3}. After that, in Sec. \ref{sec4}, the characteristics of the phenomenon under different system parameters, including the built-in mechanical frequency, mechanical quality factor, and especially the respective powers of the drive tones, are discussed in details. Subsequently, the variations of the oscillatory staircase phenomenon, when the frequencies of two drive tones are mismatched, are analyzed in Sec. \ref{sec5}, before the session of discussion and conclusion.

	\section{Setup and associated dynamics}
	\label{sec2}
	
The COMS considered here can be represented by a Fabry-P\`erot cavity with a movable mirror, which acts as a mechanical resonator driven by the intracavity field. The dynamics of the system is modeled by the coupled intracavity and mechanical modes, as the following differential equations,
	\begin{align}
		\dot{a}&=-\kappa a+ig_mX_ma+E(t),\nonumber\\
		\dot{X}_m&=\omega_mP_m,\nonumber\\
		\dot{P}_m&=-\omega_mX_m-\gamma_mP_m+g_m|a|^2+\sqrt{2\gamma_m}\xi_m(t),
		\label{dy}
	\end{align}
in a reference system rotating at the cavity resonance frequency $\omega_c$,
where $g_m$ is the optomechanical coupling strength at the single-photon level, $\gamma_m$ the mechanical damping rate, and $X_m=\sqrt{m\omega_m/\hbar}\times x_m$ ($m$ is 
the effective mass of mechanical resonator) and $P_m$ represent the dimensionless displacement and momentum of the mechanical resonator, respectively. In addition to the external coupling rate $\kappa_e$, the intrinsic loss rate $\kappa_i$ contributes to the total cavity damping rate $\kappa=\kappa_e+\kappa_i$. While the effect of cavity vacuum noise can be neglected, the thermal Langevin noise affects the motion of the mechanical resonator, satisfying the correlation relation $\langle \xi_m(t)\xi_m(t')\rangle=(2n_{th}+1)\delta(t-t')$, where the thermal phonon occupancy is given by $n_{th}=1/(e^{\hbar \omega_m/k_BT}-1)$ \cite{noise}.

\begin{figure}[tb]
	\centering\includegraphics[width=8cm]{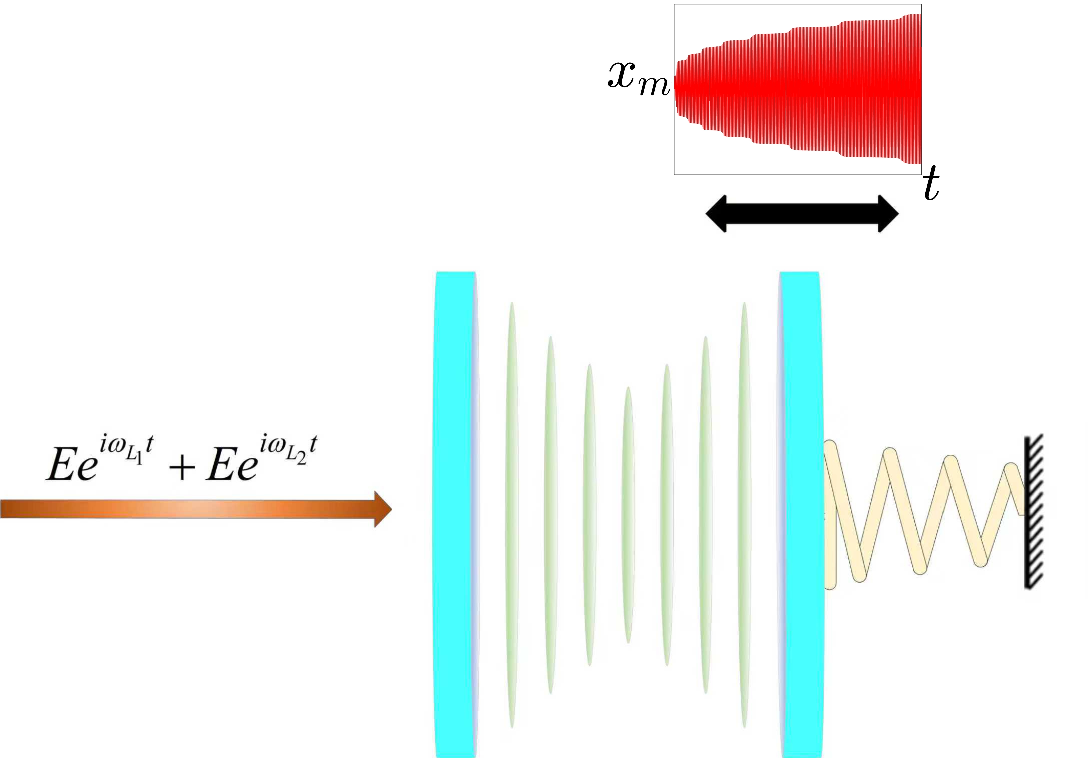}
	\caption{An OMS driven by a two-tone field with the example of Fabry-P\`erot (FP) cavity with movable mirror. If the two tones of driving laser field satisfy Eq. (\ref{fm}), the displacement of mechanical resonator (the movable mirror) can evolve through a number of quasi-stable states of step like as in the inset. To have the variable step sizes, the ratio $\omega_m/\kappa$ should be within a certain regime. We demonstrate the step wise evolution by means of the associated mechanical energy in the other figures, so that the feature is even clearer. }  \label{fig1}
\end{figure}

\begin{figure*}[tbh]
	\centering\includegraphics[width=17.5cm]{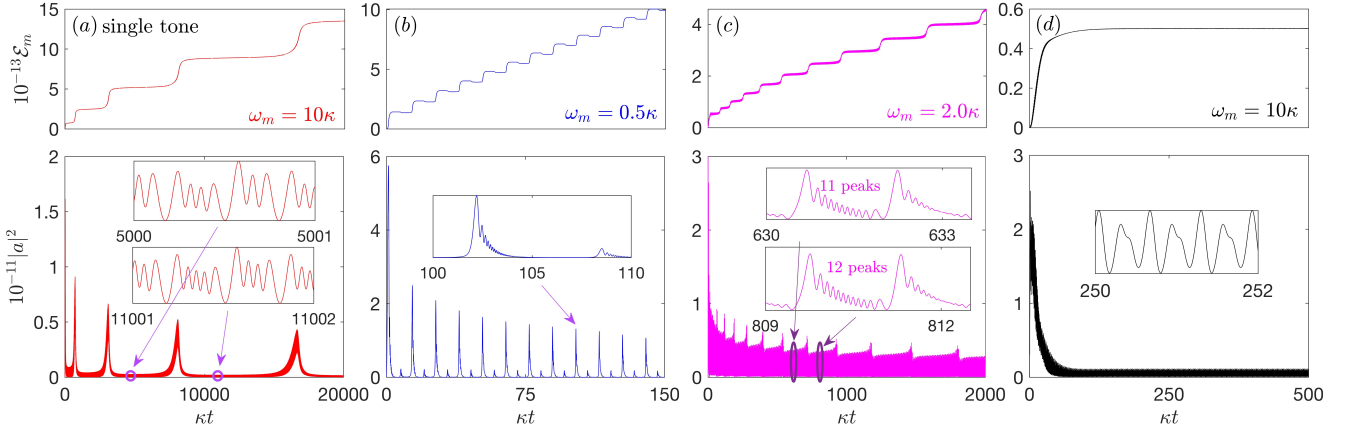}
	\caption{Four different scenarios of evolving mechanical energy and intracavity field intensity. (a) A COMS in the resolved sideband regime ($\omega_m/\kappa=10$) is driven by a single-tone field blue-detuned at $\Delta=-\omega_m$. The inset shows that the patterns of the continuous intracavity field in those quasi-stable states are similar but have the increased peak numbers from one step to another. (b) A COMS in the unresolved sideband regime ($\omega_m/\kappa=0.5$) is driven by a two-tone field satisfying Eq. (\ref{fm}). Here, only pulsed intracavity field exists around each termination of half mechanical oscillation period and one mechanical oscillation period. The prominent pulses after each mechanical oscillation push the mechanical resonator to a higher energy step. (c) A COMS in the intermediate regime ($\omega_m/\kappa=2$) is driven by a two-tone field satisfying Eq. (\ref{fm}). In this scenario there exist both continuous and pulsed intracavity field 
		at different intervals of time. The continuous ones corresponding to each mechanical energy step also have a pattern of regular development (an extra field peak number is added after the mechanical energy goes up to a higher step); see the insets. (d) A COMS in the resolved sideband regime ($\omega_m/\kappa=10$) is driven by a two-tone field satisfying Eq. (\ref{fm}). In this scenario both 
		intracavity field and mechanical resonator quickly stabilize to a locked state \cite{lo1, lo2}. The fixed parameters 
		are given as $g_m=10^{-5}\kappa$, $\gamma_m=0$. The drive amplitude of the single tone drive is $E=9.0\times10^5\kappa$. All system parameters are scaled by the cavity damping rate $\kappa$ in the numerical calculations and, correspondingly, the evolution time is scaled to the dimensionless one $\kappa t$. The two-tone drive with the respective detunings $\Delta_1=0$ and $\Delta_2=-\omega_m$ has $\sqrt{2}E=9.0\times10^5\kappa$ so that the total power is the same as that of the single-tone drive in (a).}  \label{fig2}
\end{figure*}

In the presence of optomechanical nonlinearity, the system driven by a single-tone field 
$E(t)=Ee^{i\Delta t}$, where $E=\sqrt{2\kappa_e P/\hbar\omega_{l}}$ ($P$ is the driving laser power and $\Delta=\omega_c-\omega_l$ is the detuning of the drive frequency from the cavity resonance frequency), undergoes a transition from static steady state to  oscillatory stability, once the drive power is above a threshold. It is known as a Hopf bifurcation \cite{OMS, ss1, ss2}. In this scenario, the stabilized mechanical motion
\begin{align}
	X_m=A_m\cos(\Omega_m t)+d_m, \label{xmt}
\end{align}
where $A_m$ is the amplitude of the oscillation and $d_m$ a static displacement, corresponds to the intracavity field intensity
\begin{align}
	|a(t)|^2=\sum_{l=0}^{+\infty}A_l\cos(l\cdot\Omega_mt+\psi_l),
	\label{f-s}
\end{align}
where the observed mechanical frequency $\Omega_m=\omega_m+\delta_{os}$ can differ from the built-in mechanical frequency $\omega_m$ due to the optical spring effect. Then the first sideband $A_1$ of the field intensity predominantly determines 
the stabilized mechanical energy 
\begin{align}
	\langle\mathcal{E}_{m,st}\rangle=\frac{(g_m\omega_mA_1)^2}{2\delta^2_{os}(2\omega_m+\delta_{os})^2+2\gamma_m^2(\omega_m+\delta_{os})^2}, 
	\label{m-e}
\end{align}
since its frequency is the closest to the mechanical resonance point. 
   
Near the blue-detuned point $\Delta=-\omega_m$, the system driven by a single-tone field will have a peculiar behavior of stabilization in an anomalous way within a certain range of driving powers \cite{as}. As shown in Fig. \ref{fig2}(a), in this regime the mechanical energy rises in a step-like manner; the mechanical resonator and the intracavity field go through a series of metastable states. Between two metastable states, the field intensity is totally different from the one in Eq. (\ref{f-s}) 
but in a pulsed form of drastic change, so that it successively pushes the mechanical resonator to different orbits until the system reaches the final stabilization determined by the fabrication parameters. The intracavity field intensity evolves in a regular pattern to have the peak number increased by one whenever the mechanical energy goes up to a higher step; see the inset in Fig. \ref{fig2}(a), while the observed mechanical oscillation frequency also varies in these different metastable states. As the result, the stabilized energy values due to the ever-changed first sideband amplitude $A_1$ and mechanical frequency shift $\delta_{os}$ on each step, which are determined according to Eq. (\ref{m-e}), well match the observed mechanical energy values on the different steps in Fig. \ref{fig2}(a). We will show later that after the driving field is replaced a special two-tone one, the possible step-like evolution will become much more complicated to lose this property. 

\begin{figure*}[t]
	\centering\includegraphics[width=14cm]{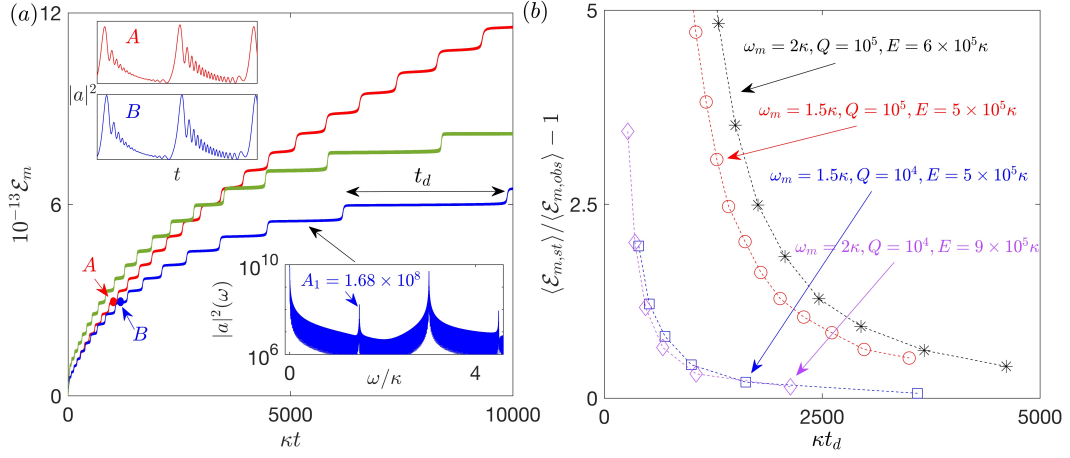}
	\caption{(a) Comparison of the evolution trajectories for the mechanical resonators of the same built-in frequency $\omega_m=1.5\kappa$, but with the different quality factors: $Q=10^5$ for the red curve and $Q=10^4$ for the blue curve. 
		The other parameters for the curves are indicated by those of the corresponding colors in (b). The other one with $Q=10^4$, the green curve, is due to a drive amplitude $E=6\times10^5\kappa$ higher than those of the blue and green curve under $E=5\times10^5\kappa$. Here two drives tones are fixed at $\Delta_1=0$, $\Delta_2=-\omega_m$. On the same level of steps A and B counted from the starting point, the intracavity field patterns are identical. The first-order sideband for the continuous intracavity fields between the pulses can be found by their Fourier transform, as the example in the lower inset. (b) The relations between the step duration $t_d$ and the discrepancy between the observed mechanical energy and the stabilized mechanical energy due to the action of the first-order sideband of the continuous intracavity field, which lasts between every two pulsed ones. These two values of energy will tend to be consistent with the increased duration or increased size of the displayed energy step.}  \label{fig3}
\end{figure*}

\section{Staircase evolution patterns induced by two-tone drives}
\label{sec3}

Now we consider different scenarios that the system is driven by a two-tone laser field $E(t)=\sum_{j=1}^2E_je^{i\Delta_j t}$, 
and the two-tone field satisfies the following frequency-matching condition.
\begin{align}
|\omega_{l1}-\omega_{l2}|=|\Delta_1-\Delta_2|=\omega_m. \label{fm}
\end{align}
This setting can lead to rich dynamical behaviors of COMS, since optomechanical nonlinearity can be activated much easier in this way. The associated dynamical behaviors of the setting are closely relevant to the ratio $\omega_m/\kappa$, as we illustrate in what follows.

The first scenario is in the unresolved sideband regime with $\omega_m<\kappa$. It is seen from Fig. \ref{fig2}(b) that the mechanical energy of the system increases in a manner of highly regular step. This phenomenon is known as a nonlinear resonance phenomenon \cite{nr}. Different from the anomalous stabilization process due to a single-tone drive, each mechanical energy step in this case lasts over one exact oscillation period $2\pi/\omega_m$, and the intracavity field mostly exists as the pulses that successively pushes the mechanical resonator to the higher orbits. Within each oscillation period, the mechanical resonator mainly moves with its inertia except for receiving a backward force from a tiny pulse at each half oscillation period. In the frequency-domain, these pulsed fields consist of equally spaced sidebands, whose number increases rapidly with the ascending steps and can thus be used to broaden their bandwidth in an application of optical frequency combs \cite{ofc}. On the other hand, when the mechanical frequency is much higher than the cavity damping rate, i.e. $\omega_m\gg \kappa$ as in the resolved sideband regime, there will be another scenario of the locking of mechanical motion into the discrete orbits, whose mechanical oscillation amplitudes can become independent of the drive power \cite{lo1, lo2}, though the associated time-evolution can be as simple as the example in Fig. \ref{fig2}(d). In other words, within the same mechanical orbit, the mechanical amplitude, possibly together 
with the mechanical oscillation phase, becomes totally frozen without increasing with the driving power.  

There exists another important dynamical pattern in the intermediate regime where the built-in mechanical frequency is not so larger than the cavity damping rate, e.g., $\omega_m=2\kappa$. It exhibits the evolving mechanical energy in a form as the one in Fig. \ref{fig2}(c), where the energy growth is step wise but has the varied step sizes. This scenario looks like an anomalous stabilization process \cite{as} but has its very different properties.

One of its features is that the development of orbital steps is mainly determined by the built-in mechanical frequency $\omega_m$ and the quality factor $Q=\omega_m/\gamma_m$ of mechanical resonator, but is less relevant to the drive power. As shown in Fig. \ref{fig3}, with the different quality factors $Q$, the systems will evolve to very different mechanical energy levels under the same pump drive power, and a higher pump power may not drive the resonator to the higher levels; compare the evolution trajectories in Fig. \ref{fig3}(a). 
Another feature is about the real-time intracavity field patterns on the mechanical energy steps. To the systems of different parameter sets, one sees that those field patterns on the same number of steps (counted from the beginning of dynamical evolution) are exactly same. One example is with the intervals A and B, respectively on the blue and red curve of two different systems in Fig. 3(a). Both of them are on the tenth step from the beginning of dynamical evolution, and they have the identical field oscillation pattern shown in the insets. Some properties of this dynamical process are similar to those of anomalous stabilization under a single-tone drive \cite{as}. However, under the same driving power, each energy step has a shorter duration and rises more rapidly. While the mechanical frequency $\omega_m$ plays an important role in determining various properties, the mechanical quality factor and driving power can jointly affect the duration $t_d$ on each orbital step. A higher quality factor leads to a more rapid increase in the mechanical energy, thus reducing the duration $t_d$ on each orbit. The staircase evolution will last forever in the ideal situation of vanishing mechanical damping $\gamma_m=0$.

Another difference from the patterns in Figs. \ref{fig2}(a) and \ref{fig2}(b) is that the mechanical energy is not close to a constant but is oscillatory on each orbital step. Due to the mechanical frequency locking under a two-tone drive satisfying the condition in Eq. (\ref{fm}), the mechanical motion on each energy step in Fig. \ref{fig2}(c) can be approximated as $X_m(t)\approx A_m\cos(\omega_m t)+d_m$ (the amplitude $A_m$ changes insignificantly during the evolution through one typical orbital step), corresponding to the observed mechanical energy
\begin{align}
\mathcal{E}_{m,obs}(t)=\frac{1}{2}A_m^2+A_md_m\cos(\omega_m t). \label{o-e}
\end{align}
Due to a larger static displacement $d_m$ for a mechanical resonator of less frequency $\omega_m$, which can be displaced 
more significantly under a less elastic restoring force, the mechanical energy has a higher amplitude $A_md_m$ to see 
its more obvious oscillatory behavior. 

Plugging the above-mentioned $X_m(t)$ into the dynamical equations, Eq. (\ref{dy}), with $E(t)=\sum_{j=1}^2E_je^{i\Delta_j t}$, one will obtain the intracavity field   
\begin{align}
	a(t)&\approx e^{i\phi(t)}\sum_{l=-\infty}^{+\infty}\left[\alpha_{l,1}e^{i(l\cdot\omega_m+\Delta_1) t}+\alpha_{l,2}e^{i(l\cdot\omega_m+\Delta_2) t}\right],\nonumber\\
	&=e^{i\phi(t)}\sum_{l=-\infty}^{+\infty}(\alpha_{l,1}+\alpha_{l+1,2})e^{i(l\cdot\omega_m+\Delta_1) t},\nonumber\\
	&=\sum_{l=-\infty}^{+\infty}\beta_le^{i(l\cdot\omega_m+\Delta_1) t+i\phi(t)}
	\label{fs}
\end{align}
under the two-tone drive, where $l$ is an integer, $\phi(t)=g_mA_m/\omega_m\sin(\omega_mt)$, and
\begin{align}
	\alpha_{l,1(2)}&=\frac{E}{\kappa}\frac{J_l(-g_mA_m/\omega_m)}{il\cdot\omega_m/\kappa+1-i(g_md_m-\Delta_{1(2)})/\kappa},
	\label{field}
\end{align}
with $J_l(x)$ being the $l$th-order Bessel function of the first kind. The corresponding intracavity field intensity or intracavity photon number takes the form,	
\begin{align}
	|a(t)|^2=\sum_{l=0}^{+\infty}A_l\cos(l\cdot\omega_mt+\psi_l),
	\label{field2}
\end{align}
to have the sideband magnitude $A_1=2|\sum_{l=-\infty}^{+\infty}\beta^*_l\beta_{l+1}|$ on the first order. The existence of such continuous intracavity field between the pulsed ones is different from what happens in the unresolved sideband regime with only pulsed intracavity fields \cite{nr}. As in one inset of Fig. 3(a), the amplitude $A_1$ of the first-order sideband can be numerically read from the Fourier transform of the evolved continuous intracavity field intensity $|a(t)|^2$, which is directly found from the numerical simulation based on Eq. (\ref{dy}).
If the mechanical motion and intracavity field are completely stabilized, the averaged mechanical energy determined by this first-order sideband will be 
\begin{align}
	\langle\mathcal{E}_{m,st}\rangle=\frac{(g_mA_1)^2}{2\gamma_m^2}, \label{s-e}
\end{align} 
which is under the mechanical resonance due to the mechanical frequency locking to the built-in value $\omega_m$.

\begin{figure*}[t]
	\centering\includegraphics[width=13cm]{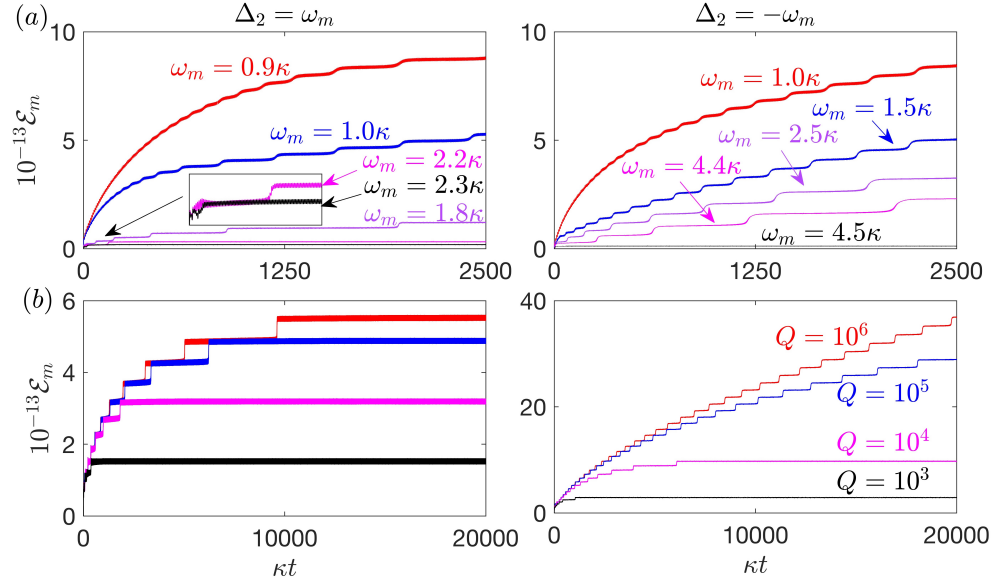}
	\caption{(a) The oscillatory staircase evolution due to different built-in mechanical frequency. Here, the fixed system parameters are $E=5\times10^5\kappa$ and $Q=10^5$. (b) The oscillatory staircase phenomenon for different mechanical quality factors $Q=\omega_m/\gamma_m$. Here we fix the parameters at $\omega_m=2\kappa$ and $E=9\times10^5\kappa$.}  \label{fig4}
\end{figure*}	

On the energy steps during dynamical evolution, the above energy determined by the first-order field intensity sideband can obviously differ from the one averaged from Eq. (\ref{o-e}), i.e. the observed mechanical amplitude $A_m$ on the quasi-stable orbits is not the same as the stabilized amplitude $g_mA_1/\gamma_m$ driven by the first-order field sideband $A_1$. For instance, by performing spectral analysis on the cavity field corresponding to one step along the blue curve in Fig. 3(a), we obtain the associated first-order sideband amplitude as $A_1=1.68\times10^8$, and the calculated average mechanical energy is $\langle\mathcal{E}_{m,st}\rangle=6.27\times10^{13}$ (the dimensionless energy is equivalent to phonon number), which is higher than the observed mechanical energy $\langle\mathcal{E}_{m,obs}\rangle=5.51\times10^{13}$ on that step. In fact, during the early stage of system evolution, the calculated mechanical energy according to Eq. (\ref{s-e}) is much higher than the actual value, even by $2\sim3$ orders of magnitude. This discrepancy is in sharp contrast to an anomalous stabilization under a single-tone drive \cite{as}, where the first-order sideband amplitude $A_1$ keeps adjusting with the mechanical frequency shift $\delta_{os}$ induced by the optical spring effect, so that the energy calculated with Eq. (\ref{m-e}) is always consistent with the observed mechanical energy. Due to the mechanical frequency locking under a two-tone drive satisfying the condition in Eq. (\ref{fm}), no optical spring effect exists, so we always have $\delta_{os}=0$ in that scenario. The system thus loses a factor contributing to its stabilization, and it explains why the energy steps in the early stage of system evolution grow so fast and why the durations $t_d$ of the quasi-stable states are much shorter than those in an anomalous stabilization under single-tone drive. On the energy steps as the quasi-stable states, the mechanical motion in the concerned intermediate regime is actually not completely stabilized, giving rise to the inequality $A_m\neq g_mA_1/\gamma_m$. This discrepancy will gradually disappear as the system gets closer to the final stabilization. Figure \ref{fig3}(b) demonstrates how the calculated energy with Eq. (\ref{s-e}) will tend to the observed mechanical energy, as the durations $t_d$ of the steps keep increasing; on the steps of longer periods the two energy values become more consistent. Such tendency toward the consistency becomes quicker to manifest for the systems with higher mechanical damping rate $\gamma_m$ or lower mechanical quality factor $Q$; see the comparison in Fig. \ref{fig3}(b). From the view of applications, such instability for COMS in the intermediate regime of mechanical frequency provides a possibility to realize more significant mechanical oscillation of narrow bandwidth (phonon laser) \cite{phonon01,phonon1, phonon11, phonon2, phonon3, phonon33,pl}, because the system with a fixed mechanical quality factor can reach a higher mechanical amplitude than the corresponding one due to the anomalous stabilization under a single-tone drive \cite{as}. It is therefore meaningful to study this particular dynamical scenario in the intermediate regime of built-in mechanical frequency.

\begin{figure*}[t]
	\centering\includegraphics[width=13cm]{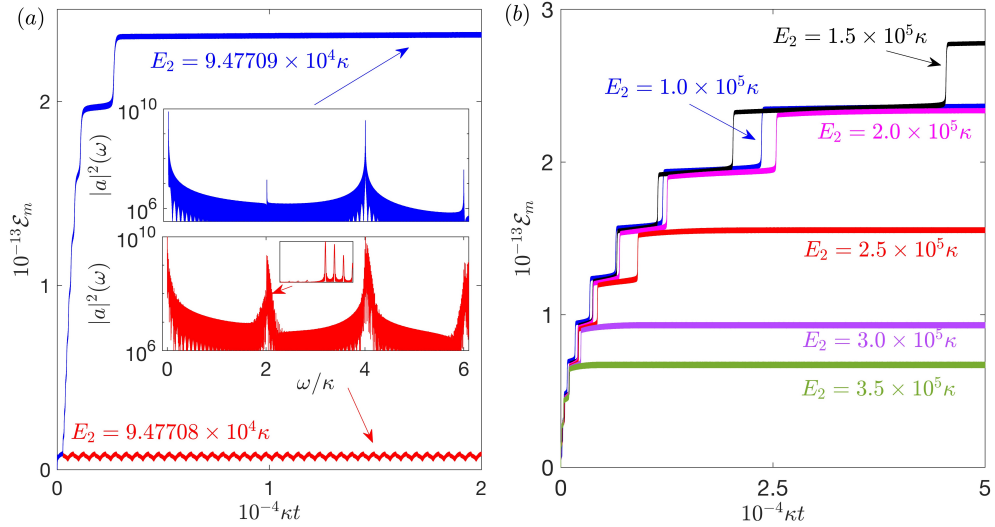}
	\caption{(a) A bifurcation due to the difference in the powers of the drive tones. Here one of the tone is fixed 
		at $E_1=5\times10^5\kappa$, and the other one is increased gradually. There is a sudden change of the associated intracavity field spectrum and the evolution pattern of mechanical energy, after the tone amplitude $E_2$ is adjusted across the indicated values. 
		(b) Corresponding relation between the evolved mechanical energy and the pump amplitude $E_2$, given the fixed $E_1=5\times10^5\kappa$. The other fixed parameters are $\Delta_1=0$, $\Delta_2=\omega_m$, $Q=10^5$, and $\omega_m=2\kappa$.}  \label{fig5}
\end{figure*}
	
\section{Oscillatory staircase phenomenon in different setups}
	\label{sec4}
Different system parameters affect the dynamical behaviors of the concerned COMS. Here we illustrate these properties in terms of the built-in mechanical frequency $\omega_m$, the mechanical quality factor $Q$, as well as the powers of the two drive tones. These dynamical properties are useful to the possible experimental implementation of the scenario.

\subsection{Built-in mechanical frequency and mechanical quality factor}

The built-in mechanical frequency serves as an important system parameter. As shown in Fig. \ref{fig2}, under the drive of a two-tone laser field of the identical power and with all other system parameters fixed to the same, the completely different dynamical patterns emerge due to the difference in mechanical frequency.
The oscillatory staircase phenomenon occurs in the regime where the built-in mechanical frequency is not very larger than the cavity decay rate.
In Fig. \ref{fig4}(a), we fix one of the drive tones to the detuning $\Delta_1=0$ and present the evolution processes of the mechanical energy due to different built-in mechanical frequencies. When the other drive tone is red-detuned with $\Delta_2=\omega_m$, the oscillatory staircase phenomenon will disappear and transit to the locked orbits \cite{lo1, lo2} after the mechanical frequency increases to $\omega_m=2.3\kappa$. On the other hand, if the mechanical frequency decreases to $\omega_m=0.9\kappa$, the distinction between the individual steps gradually becomes blurred, and the oscillatory staircase phenomenon will vanish as well.

\begin{figure*}[t]
	\centering\includegraphics[width=17cm]{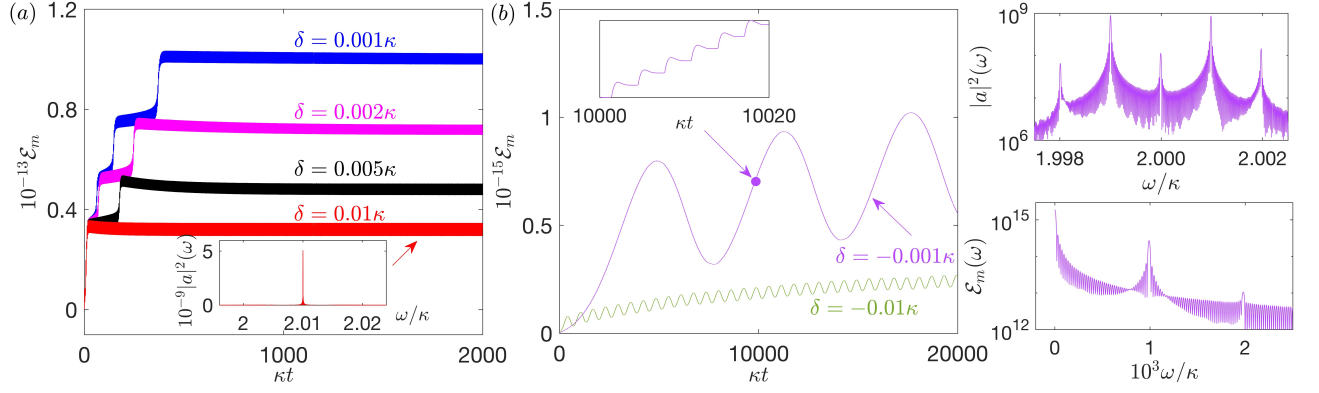}
	\caption{(a) The dynamical processes due to the positive values of frequency mismatch $\delta$. The oscillatory staircase phenomenon gradually disappears as the mismatch $\delta$ increases. Here the field sidebands are shifted by an amount of the frequency mismatch; see the inset. (b) The dynamical processes due to the negative values of frequency mismatch $\delta$. A long-period oscillation wit the period $2\pi/\delta$ manifests for the evolving mechanical energy. Here the field intensity sidebands are significantly modified by the drive frequency mismatch: a large number of newly generated sidebands, with their intervals of $n\delta$ ($n$ are the integers) from the original ones (due to the frequency matched two drive tones) at $l\cdot \omega_m$ ($l$ are integers). For instance, as seen from the top of the right frame, the newly generated field sidebands distanced by the frequency mismatch $\delta=10^{-3}\kappa$ to the one at the point $\omega=\omega_m=2\kappa$ have much larger amplitudes, leading to the mechanical energy spectrum in the lower part, where the oscillation components are also distanced by the interval $\delta=10^{-3}\kappa$. In all these figures, we set $\Delta_1=0$ and $\Delta_2=-\omega_m$ (this tone is blue-detuned instead of the red-detuned one in Fig. 5), and $E_1=E_2=5\times 10^5\kappa$.}  \label{fig6}
\end{figure*}

One can also set the other tone to blue-detuned at $\Delta_2=-\omega_m$. Then the oscillatory staircase phenomenon will be converted to the locked orbits with a higher mechanical frequency at $\omega_m=4.5\kappa$; see the right frame of Fig. \ref{fig4}(a). Apparently, the mechanical frequency range manifesting the oscillatory staircases becomes wider in this case. The combination of two tones with $\Delta_1=0$ 
and $\Delta_2=-\omega_m$ is more flexible for having the oscillatory staircase phenomenon.

With respect to the mechanical quality factor $Q=\omega_m/\gamma_m$, a higher quality factor enables the system to stabilize at a higher final orbital step. As illustrated in Fig. \ref{fig4}(b), after the quality factor is decreased from $Q=10^6$ to $Q=10^3$, the oscillatory staircase phenomenon will disappear for the drive tone combination $\Delta_1=0$ 
and $\Delta_2=-\omega_m$, while only a remnant oscillatory staircase exists for the other combination $\Delta_1=0$ 
and $\Delta_2=\omega_m$. Obviously, with one of the tones being $\Delta_2=-\omega_m$ of blue-detuned, the system will evidently have much more orbital steps and stabilize at a much higher orbital step.

\subsection{Unequal drive tone powers}

 When the powers of the two drive tones are different, the dynamical evolution of the concerned system will have an interesting feature. As seen from Fig. \ref{fig5}, we study an example with the drive amplitude of one tone with $\Delta_1=0$ being fixed at $E_1=5\times10^5\kappa$, while the other tone with $\Delta_2=\omega_m$ varies its power over a range. For a smaller drive amplitude, e.g., $E_2\leqslant9.47708\times10^4\kappa$, the mechanical resonator exhibits a complex oscillation pattern without having a staircase evolution behavior, as what is represented by the red curve in Fig. \ref{fig5}(a). The complex pattern of oscillation is seen from the corresponding intracavity field spectrum (the red colored one in an inset): besides the main sideband peaks at $l\cdot\Omega_m$ ($l$ are the integers), the equally spaced sub-sidebands are excited around the main peaks.
In this situation the tone with the detuning $\Delta_1=0$ dominates, so that 
the optical spring effect can exist to lose the locking of mechanical frequency and lead to a small mechanical frequency shift $\delta_{os}$. The generated small sub-sidebands have the interval of the shift $\delta_{os}$. Once the drive power of the blue-detuned tone exceeds a threshold, e.g., $E=9.47709\times10^4\kappa$ in Fig. \ref{fig5}(a), there will be a sudden transition to the staircase evolution pattern to eliminate the sub-sidebands; see the other blue colored inset in Fig. \ref{fig5}(a). The mechanical energy is suddenly enhanced across this point of special bifurcation induced by the power difference in the drive tones. 

As the power of the drive tone with $\Delta_2=\omega_m$ continues to increase, for instance, to $E_2=1.0\times10^5\kappa$ and $E_2=1.5\times10^5\kappa$, the final stable step level of mechanical energy will correspondingly 
rise as in Fig. \ref{fig5}(b). However, when the power of the tone with $\Delta_2=\omega_m$ is enhanced further, for example, to $E_2=2.0\times10^5\kappa$, the final stable step level will decrease instead. This phenomenon is similar to the anomalous stabilization phenomenon under a single tone drive \cite{as}, where the best enhancement of mechanical energy also occurs at an optimum drive power in the mid of feasible power range. Choosing the appropriate drive power can, therefore, lead to a high mechanical orbit.
	
\section{Effects of mismatched drive tones}	
\label{sec5}

In practical situations, there could exist the imperfect frequency match for the drive tones to deviate from the condition in Eq. (\ref{fm}), so that one has $|\Delta_1-\Delta_2|=\omega_m+\delta$. As it is expected, a large mismatch $\delta$ will destroy 
the oscillatory staircase evolution phenomenon. However, such imperfect scenarios also have their interesting dynamical properties, which are determined by the sign of the frequency shift $\delta$. In Fig. \ref{fig6}(a), we show that when the frequency shift is positive $\delta>0$, the step wise evolution phenomenon will gradually disappears as the shift $\delta$ increases. The corresponding spectrum of the intracavity field intensity also indicates that the position of the first-order sideband gradually shifts away from the built-in mechanical frequency $\omega_m$ to $\omega_m+\delta$ (here the shift $\delta_{os}$ induced by optical spring effect is neglected due to fact $\delta_{os}\ll\delta$). To maintain a clearer process of oscillatory staircase evolution, one should reduce the positive frequency mismatch $\delta$ to less than $0.01\kappa$.	
	
On the other hand, when the mismatch $\delta$ becomes negative, the totally different dynamical behaviors of the system will be much more complex. As shown in the inset of the left frame of Fig. \ref{fig6}(b), the mechanical energy increases in a pattern of smooth steps within a short period of time $T_S=2\pi/\omega_m$, similar to the energy steps generated through a nonlinear resonance \cite{nr}. However, there exists a long-period oscillatory pattern with its period $T_L=2\pi/\delta$. The cause of this phenomenon is the excitation of a large number of field sidebands around the points $\omega=n \omega_m$ ($n$ are the integers), where the sidebands under the drive frequency match locate. These newly generated sideband even overshadow the original one by amplitude; see the top part in the right frame of Fig. 6(b). Due to such significantly modified field sidebands, the mechanical motion will become highly complicated with the evolving energy
\begin{align}
	\mathcal{E}_m(t)\approx &E_0+E_{D,1}\cos(\delta t)+\cdots+E_m\cos(\omega_mt)+\cdots,
	\label{mec}
\end{align}
where $E_m\ll E_{D,1}, E_0$. In other words, the oscillation component at the built-in mechanical frequency $\omega_m$ has much lower amplitude than many other components, especially the one with the long period  $T_L=2\pi/\delta$, which arises from a beat frequency of the field intensity $|a(t)|^2$.  

The relevance of dynamical behaviors to the sign of frequency mismatch from the condition in Eq. (\ref{fm}) is unique in the intermediate regime of mechanical frequency $\omega_m$. This character can have important applications. More specifically, after fixing the frequency difference of two drive tones, the loss of staircase evolution pattern due to the mismatch $\delta$ can be detected by the field spectrum. It can be used to deduce the mass of an added tiny particle to the mechanical resonator.

	\section{Discussion and conclusion}
	\label{conclusion}	
	In the numerical simulations of the concerned dynamical processes, we adopt the system parameters scaled with the total damping rate $\kappa$ of the used optical cavity. If we consider a range of $\kappa$ between $1$ MHz and $10$ MHz, which is consistent with those of suspended mechanical membranes \cite{sus} and high-quality silica optical microresonators \cite{pl,mc1,mc2}, the required powers for reaching a typical drive amplitude $E=5\times 10^5\kappa$ for seeing the phenomenon will be $32.4$ mW for $\kappa=1$ MHz and $324$ mW for $\kappa=10$ MHz (given the pump field wavelength at $1537$ nm). After the used optomechanical coupling constant is enhanced from $g_m=10^{-5}\kappa$ to $10^{-4}\kappa$, these powers can be reduced by $100$ times. Based on the evaluation, it is possible to observe the oscillatory staircase evolution in various platforms.
	
	We have presented a study of oscillatory staircase evolution phenomenon, which occurs in a two-tone driven COMS with its built-in mechanical frequency at the intermediate values not so less and not so larger than the associated cavity damping rate. This scenario has very different properties form those of anomalous stabilization under single-tone drive \cite{as}, as well as from those of nonlinear resonance in the unresolved sideband regime and under a two-tone drive \cite{nr}. It exhibits an incomplete stabilization in the quasi-stable states and is more sensitive to the fabricated mechanical parameters (the mechanical frequency $\omega_m$ and mechanical quality factor $Q$). Under the same drive power, the systems with varied $\omega_m$ and $Q$ can evolve to very different optomechanical oscillations. With the unmatched tone powers and tone frequencies, a COMS in the concerned regime respectively demonstrates a special bifurcation determined by the drive tone power difference and the totally different dynamical behaviors determined by the sign of tone frequency deviation. These special properties only exist in the systems built with the mechanical frequency within the intermediate range. The dynamical properties obtained through the current research can lead to the potential applications in phonon laser generation and high-precision sensing of mass and other quantities.
	
	\begin{acknowledgments}
		This work was supported by National Natural Science Foundation of China (Grant No. 12374348), Natural Science Foundation of Fujian Province (Grant No. 2024J01078), and ANID Fondecyt
		Regular (Grant No. 1221250). 
	\end{acknowledgments}


\begin{thebibliography}{99}
\bibitem{OMS} M. Aspelmeyer, T. J. Kippenberg, and F. Marquardt, Cavity optomechanics, Rev. Mod. Phys. 86, 1391
	    (2014).
\bibitem{OMIT1}G. S. Agarwal and S. Huang, Electromagnetically induced transparency in mechanical effects of light, Phys. Rev. A 81, 041803 (2010).
\bibitem{OMIT2}S. Weis, R. Rivi\`ere, S. Del\'eglise, E. Gavartin, O. Arcizet, A. Schliesser, and T. J. Kippenberg, Optomechanically Induced Transparency, Science 330, 1520 (2010).
\bibitem{OMIT3}A. H. Safavi-Naeini, T. P. Mayer Alegre, J. Chan, M. Eichenfield, M. Winger, Q. Lin, J. T. Hill, D. E. Chang, and O. Painter, Electromagnetically induced transparency and slow light with optomechanics, Nature 472, 69 (2011).
\bibitem{OMIA1}F. Massel, T. T. Heikkil\"a, J.-M. Pirkkalainen, S. U. Cho, H. Saloniemi, P. J. Hakonen, and M. A. Sillanp\"a\"a, Microwave amplification with nanomechanical resonators, Nature 480, 351 (2011).
\bibitem{OMIA2} A. Nunnenkamp, V. Sudhir, A. K. Feofanov, A. Roulet, and T. J. Kippenberg, Quantum-limited amplification and parametric instability in the reversed dissipation regime of cavity optomechanics, Phys. Rev. Lett. 113, 023604 (2014).
\bibitem{OMIA3}A. Metelmann and A. A. Clerk, Quantum-limited amplification via reservoir engineering, Phys. Rev. Lett. 112, 133904 (2014).

\bibitem{nonre1}M. Hafezi and P. Rabl, Optomechanically induced nonreciprocity in microring resonators, Opt. Express 20, 7672 (2012).
\bibitem{nonre2}Z. Shen, Y.-L. Zhang, Y. Chen, C.-L. Zou, Y.-F. Xiao, X.-B. Zou, F.-W. Sun, G.-C. Guo and C.-H. Dong, Experimental realization of optomechanically induced non-reciprocity, Nat. Photon. 10, 657 (2016).

\bibitem{phonon01} I. S. Grudinin, H. Lee, O. Painter, and K. J. Vahala,
Phonon laser action in a tunable two-level system, Phys.
Rev. Lett. 104, 083901 (2010).

\bibitem{phonon1} H. Jing, S. K. \"{O}zdemir, X. Y. L\"u, J. Zhang, L. Yang, and F. Nori, PT -symmetric phonon laser, Phys. Rev. Lett. 113, 053604 (2014).	

\bibitem{phonon11}B. He, L. Yang, and M. Xiao, Dynamical phonon laser
in coupled active-passive microresonators, Phys. Rev. A
94, 031802(R) (2016).  

  
\bibitem{phonon2}G. Z. Wang, M. M. Zhao, Y. C. Qin, Z. Q. Yin, X. S. Jiang, and M. Xiao, Demonstration of an ultra-low-threshold phonon laser with coupled microtoroid resonators in vacuum, Photonics Res. 5, 73 (2017).
\bibitem{phonon3} J. Zhang, B. Peng, S. K. \"{O}zdemir, K. Pichler, D. O. Krimer, G. P. Zhao, F. Nori, Y. X. Liu, S. Rotter, and L. Yang, A phonon laser operating at an exceptional point, Nat. Photon. 12, 479 (2018).

\bibitem{phonon33} Y. F. Xie, Z. Cao, B. He, and Q. Lin, PT-symmetric
phonon laser under gain saturation effect, Opt. Express
28, 22580 (2020).

\bibitem{pl} X. Li, Y. Xie, S. Ding, M. Zhang, Y. He, B. He, M. Xiao, and X. Jiang, Observation of Kerr soliton microcomb locked with a phonon laser, Sci. Adv. 12, eaeb3400 (2026).		

\bibitem{ofc1} M.-A. Miri, G. D'Aguanno, and A. Al\`u, Optomechanical frequency combs, New J. Phys. 20, 043013 (2018).
\bibitem{ofc2} Y. Hu, S. L. Ding, Y. C. Qin, J. X. Gu, W. J. Wan, M. Xiao, and X. S. Jiang, Generation of optical frequency
comb via giant optomechanical oscillation, Phys. Rev. Lett. 127, 134301 (2021).

\bibitem{ofc22} R. C. Ng, P. Nizet, D. Navarro-Urrios, G. Arregui, M. Albrechtsen, P. D. García, S. Stobbe, C. M. Sotomayor-Torres, and G. Madiot,
Intermodulation of optical frequency combs in a multimode optomechanical system, Phys. Rev. Research 5, L032028 (2023).

\bibitem{ofc3}Y. Wang, M. Zhang, Z. Shen, G.-T. Xu, R. Niu, F.-W. Sun, G.-C. Guo, and C.-H. Dong, Optomechanical frequency comb based on multiple nonlinear dynamics, Phys. Rev. Lett. 132, 163603 (2024).
\bibitem{ofc} X. Gu, J. Zhang, S. Ding, X. Jiang, B. He, and Q. Lin, Optical frequency comb significantly spanned to broadband by an optomechanical resonance, Photon. Res. 12, 1981 (2024).
\bibitem{comb-s} S. Ding, B. He, Y. Wu, Y. Hu, H. Wang, W. Wan, M. Xiao, and X. Jiang, Bloch-band structure of cavity optomechanical oscillations, Phys. Rev. Research 7, L012059 (2025).

\bibitem{ms01}J.-J. Li and K.-D. Zhu, All-optical mass sensing with coupled mechanical resonator systems, Phys. Rep. 525, 223 (2013).
\bibitem{ms02}M. Sansa, M. Defoort, A. Brenac, M. Hermouet, L. Banniard, A. Fafin, M. Gely, C. Masselon, I. Favero, G. Jourdan, and S. Hentz, Optomechanical mass spectrometry, Nat. Commun. 11, 3781 (2020).

\bibitem{fs1} E. Gavartin, P. Verlot, and T. J. Kippenberg, A hybrid on-chip optomechanical transducer for ultrasensitive force measure- ments, Nat. Nanotechnol. 7, 509 (2012).
\bibitem{fs2}J. Moser, J. G\"uttinger, A. Eichler, M. J. Esplandiu, D. E. Liu, M. I. Dykman, and A. Bachtold, Ultrasensitive force detection with a nanotube mechanical resonator, Nat. Nanotechnol. 8, 493 (2013).
\bibitem{fs3}S. Schreppler, N. Spethmann, N. Brahms, T. Botter, M. Barrios,
and D. M. Stamper-Kurn, Optically measuring force near the standard quantum limit, Science 344,1486(2014). 
\bibitem{fs4}D. Mason, J. Chen, M. Rossi, Y. Tsaturyan, and A. Schliesser, Continuous force and displacement measurement below the standard quantum limit, Nat. Phys. 15, 745 (2019).
\bibitem{fs5}L. F. Buchmann, S. Schreppler, J. Kohler, N. Spethmann, and D. M. Stamper-Kurn, Complex squeezing and force measurement beyond the standard quantum limit, Phys. Rev. Lett. 117, 030801 (2016).

\bibitem{cooling1} I. Wilson-Rae, N. Nooshi, W. Zwerger, and T. J. Kippenberg, Theory of ground state cooling of a mechanical oscillator using dynamical backaction, Phys. Rev. Lett. 99, 093901 (2007). 
\bibitem{cooling2} F. Marquardt, J. P. Chen, A. A. Clerk, and S. M. Girvin, Quantum theory of cavity-assisted sideband cooling of mechanical motion, Phys. Rev. Lett. 99, 093902 (2007).
\bibitem{cooling3} Teufel, J. D., T. Donner, D. Li, J. W. Harlow, M. S. Allman, K. Cicak, A. J. Sirois, J. D. Whittaker, K. W. Lehnert, and R. W. Simmonds, Sideband cooling of micromechanical motion to the quantum ground state, Nature (London) 475, 359 (2011).
\bibitem{cooling4} J. Chan, T. P. Mayer Alegre, A. H. Safavi-Naeini, J. T.
Hill, A. Krause, S. Gr\"{o}blacher, M. Aspelmeyer, and O.
Painter, Laser cooling of a nanomechanical oscillator into
its quantum ground state, Nature 478, 89 (2011).
\bibitem{cooling5} B. He, L. Yang, Q. Lin, and M. Xiao, Radiation Pressure
Cooling as a Quantum Dynamical Process, Phys. Rev.
Lett. 118, 233604 (2017).
\bibitem{cooling6} C. Wang, Q. Lin, and B. He, Breaking the optomechanical
cooling limit by two drive fields on a membrane-in-the-
middle system, Phys. Rev. A 99, 023829 (2019).
\bibitem{cooling7} D. G. Lai, J. F. Huang, X. L. Yin, B. P. Hou, W. Li, D. Vitali, F. Nori, and J. Q, Liao, Nonreciprocal ground-state cooling of multiple mechanical resonators, Phy. Rev. A 102, 011502 (2020).
\bibitem{cooling8} D. G. Lai, J. Huang, B. P. Hou, F. Nori, and J. Q. Liao, Domino cooling of a coupled mechanical-resonator chain via cold-damping feedback, Phys. Rev. A 103, 063509 (2021).
\bibitem{entanglement1} D. Vitali, S. Gigan, A. Ferreira, H. R. B\"ohm, P. Tombesi, A. Guerreiro, V. Vedral, A. Zeilinger, and M. Aspelmeyer, Optomechanical entanglement between a movable mirror and a cavity field, Phys. Rev. Lett. 98, 030405 (2007).
\bibitem{entanglement2}T. A. Palomaki, J. D. Teufel, R. W. Simmonds, and K. W. Lehnert, Entangling mechanical motion with microwave fields, Science 342, 710 (2013).
\bibitem{entanglement20}Q. Lin, B. He, R. Ghobadi, and C. Simon, Fully quantum approach to optomechanical entanglement,
Phys. Rev. A 90, 022309 (2014).
\bibitem{entanglement3}R. Riedinger, A. Wallucks, I. Marinkovi\'{c}, C. L\"oschnauer, M. Aspelmeyer, S. K. Hong, and S. Gr\"oblacher, Remote quantum entanglement between two micromechanical oscillators, Nature (London) 556, 473 (2018).

\bibitem{entanglement4}C. F. Ockeloen-Korppi, E. Damsk\"agg, J. -M. Pirkkalainen, M. Asjad, A. A. Clerk, F. Massel, M. J. Woolley, and M. A. Sillanp\"a\"a, Stabilized entanglement of massive mechanical oscillators, Nature (London) 556, 478 (2018).
\bibitem{entanglement4b} Q. Lin, B. He, and M. Xiao, Entangling Two Macroscopic Mechanical Resonators at High Temperature, Phys. Rev. Applied 13, 034030 (2020)
\bibitem{entanglement5}S. Kotler, G. A. Peterson, E. Shojaee, F. Lecocq, K. Cicak, A. Kwiatkowski, S. Geller, S. Glancy, E. Knill, R. W. Simmonds, J. Aumentado, and J. D. Teufel, Direct observation of deterministic macroscopic entanglement, Science 372, 622 (2021).
\bibitem{entanglement6} L. M. de L\'epinay, C. F. Ockeloen-Korppi, M. J. Woolley, and M. A. Sillanp\"a\"a, Quantum mechanics? free subsystem with mechanical oscillators, Science 372, 625 (2021).
\bibitem{entanglement7} Y. F. Jiao, Y. L. Zuo, Y. Wang, W. Lu, J. Q. Liao, L. M. Kuang, and H. Jing, Tripartite Quantum Entanglement with Squeezed Optomechanics, Laser \& Photonics Reviews 18, 2301154 (2024).
\bibitem{ss1}T. Carmon, H. Rokhsari, L. Yang, T. J. Kippenberg, and K. J. Vahala, Temporal behavior of radiation-pressure-induced vibrations of an optical microcavity phonon mode, Phys. Rev. Lett. 94, 223902 (2005).
\bibitem{ss2} H. Rokhsari, T. J. Kippenberg, T. Carmon and K. J. Vahala, Radiation-pressure-driven micro-mechanical oscillator, Opt. Express 13, 5293 (2005). 
\bibitem{ss3} T. Carmon, H. Rokhsari, L. Yang, T. J. Kippenberg, and K. J. Vahala, Temporal Behavior of Radiation-Pressure-Induced Vibrations of an Optical Microcavity Phonon Mode, Phys. Rev. Lett. 94, 223902 (2005).
\bibitem{ss4} T. J. Kippenberg, H. Rokhsari, T. Carmon, A. Scherer, and K. J. Vahala, Analysis of Radiation-Pressure Induced Mechanical Oscillation of an Optical Microcavity, Phys. Rev. Lett. 95, 033901 (2005).
\bibitem{ss5} C. Metzger, M. Ludwig, C. Neuenhahn, A. Ortlieb, I. Favero, K. Karrai, and F. Marquardt, Self-Induced Oscillations in an Optomechanical System Driven by Bolometric Backaction, Phys. Rev. Lett. 101, 133903 (2008).
\bibitem{ss5b} S. Zaitsev, A. K. Pandey, O. Shtempluck, and E. Buks, Forced and self-excited oscillations of an optomechanical cavity, Phys. Rev. E 84, 046605 (2011).
\bibitem{ss6} J. B. Khurgin, M. W. Pruessner, T. H. Stievater, and W. S. Rabinovich, Optically pumped coherent mechanical oscillators: the laser rate equation theory and experimental verification, New J. Phys. 14, 105022 (2012).
\bibitem{ss7} M. Poot, K. Y. Fong, M. Bagheri, W. H. P. Pernice, and H. X. Tang, Backaction limits on self-sustained optomechanical oscillations, Phys. Rev. A 86, 053826 (2012).
\bibitem{ss8} O. Suchoi, L. Ella, O. Shtempluk, and E. Buks, Intermittency in an optomechanical cavity near a subcritical Hopf bifurcation, Phys. Rev. A 90, 033818 (2014).
\bibitem{ss80} M. Gao, F.-C. Lei, C.-G. Du, and G.-L. Long, Self-sustained
oscillation and dynamical multistability of optomechanical systems
in the extremely-large-amplitude regime, Phys. Rev. A 91,
013833 (2015).
\bibitem{ss9} C. Wurl, A. Alvermann, and H. Fehske, Symmetry-breaking oscillations in membrane optomechanics, Phys. Rev. A 94, 063860 (2016).
\bibitem{ss91} Q. Lin, B. He, and M. Xiao, Catastrophic transition between
dynamical patterns in a phonon laser, Phys. Rev. Res. 3,
L032018 (2021).
\bibitem{ss92} S. Christou, V. Kovanis, A. E. Giannakopoulos, and Y.
Kominis, Parametric control of self-sustained and selfmodulated
optomechanical oscillations, Phys. Rev. A
103, 053513 (2021).
\bibitem{ss10} H. Zhang, V. Eremeev, J. Wu, M. Orszag, and B. He, Scaling behaviors in optomechanically induced nonlinear oscillation, Phys. Rev. E 111, 014208 (2025).
\bibitem{as}J. L. Zhang, M. Orszag, M. Xiao, X. S. Jiang, Q. Lin, and B. He, Highly Correlated Optomechanical Oscillations Manifested by an Anomalous Stabilization, Phys. Rev. Lett. 133, 103602 (2024).
\bibitem{nr}Q. Lin, Y. Wu, G. Li, and B. He, Nonlinear optomechanical resonance entering a self-organized energy transfer pattern, Chaos Soliton. Fract. 173, 113624 (2023).
\bibitem{lo1}B. He, Q. Lin, M. Orszag, and M. Xiao, Mechanical oscillations frozen on discrete levels by two optical driving ?elds, Phys. Rev. A 102, 011503(R) (2020).
\bibitem{lo2}Y. Wu, G. Li, B. He, and Q. Lin, Amplitude and phase locking of mechanical oscillation driven by radiation pressure, Phys. Rev. A 105, 013521 (2022).

\bibitem{ms1}G. Li, Y. Wu, Y. L. Zhang, B. He and Q. Lin, Ultra-high resolution mass sensing based on an optomechanical nonlinearity, Opt. Express, 30, 15858 (2022).
\bibitem{ms2}J. W. Zheng, J. L. Zhang, Y. Z. Li, L. J. Mart\'{\i}nez, B. He, and Q. Lin, Nonlinear optomechanically induced frequency locking and its application to room temperature mass sensing, Opt. Express, 33, 31988 (2025).
\bibitem{wf1}Z. F. Yan, B. He and Q. Lin, Optomechanical force sensor operating over wide detection range, Opt. Express 31, 16535 (2023).
\bibitem{wf2}Z. F. Yan, B. He and Q. Lin, Force sensing with an optomechanical system at room temperature, Phys. Rev. A 107, 013529 (2023).

\bibitem{noise} C. W. Gardiner and P. Zoller, \textit{Quantum Noise} (Springer Verlag, 2000).
\bibitem{sus} J. Sheng, X. Wei, C. Yang, and H. Wu, Self-Organized Synchronization of Phonon Lasers, Phys. Rev. Lett. 124, 053604 (2020).
\bibitem{mc1} X. Li, K. Qi, Y. Wu, X. Wu, M. Hu, Z. Li, Y. He, S. Ding, Z. Xie, H. Zhou, B. He, M. Xiao, and X. Jiang, 
Generation of 2/3-octave-spanning visible Kerr soliton microcomb, Adv. Photon. 7, 056002 (2025).

\bibitem{mc2} Q. Shi, J. Tian, S. Ding, Y. Wang, S. Lei, M. Zhang, W. Wan, X. Ji, B. He, M. Xiao, and X. Jiang, On-chip ultra-high-Q optical microresonators approaching the material absorption limit, Photonics Res. 13, 2409 (2025).



\end{thebibliography}
\end{document}